\documentclass[aps,prb,twocolumn,amsmath,amssymb,groupedaddress]{revtex4}
\usepackage{graphics,epsfig}
\usepackage{bm}

\newcommand{\pdag}{{\phantom{\dagger}}}
\newcommand{\YbRhSi}{YbRh$_2$Si$_2$}
\newcommand{\CeAu}{CeCu$_{6-x}$Au$_x$}
\begin{document}


\title{Thermodynamic signatures of a fractionalized Fermi liquid}

\author{Andreas Hackl}
\affiliation{Department of Physics, California Institute of Technology, Pasadena, CA 91125}
\author{Ronny Thomale}
\affiliation{Department of Physics, Princeton University, Princeton, NJ 08544}
\date{\today}

\begin{abstract}
Several heavy-fermion metals display a quantum phase transition 
from an antiferromagnetic metal to a heavy Fermi liquid. In some
materials, however, recent experiments seem to find 
that the heavy Fermi liquid phase can be directly tuned into a non-Fermi liquid phase 
without apparent magnetic order.
We analyze a candidate state for this scenario 
where the local moment 
system forms a spin liquid with gapless fermionic excitations. We discuss the thermal conductivity 
and spin susceptibility of this fractionalized state both in two and,
in particular, three spatial 
dimensions for different temperature regimes. We derive a variational functional for the thermal conductivity and solve it with 
a variational ansatz dictated by Keldysh formalism.
In sufficiently clean samples and for an appropriate temperature window, we find that thermal transport is dominated by
the spinon contribution which can be detected by a characteristic maximum in the
Wiedemann-Franz ratio. For the spin susceptibility, the conduction electron Pauli paramagnetism
is much smaller than the spinon contribution whose temperature
dependence in three dimensions is logarithmically enhanced as compared to the Fermi 
liquid result.
\end{abstract}

\maketitle


\section{\label{intro} Introduction}

In many heavy fermion systems, the heavy Fermi liquid phase seems to disappear
at a zero temperature phase transition upon tuning a certain external control parameter.\cite{rmp07} 
Since the discovery of such transitions, it has remained an open problem what precisely happens
at the underlying quantum critical point (QCP). One peculiarity is 
that often, a simultaneous occurence of magnetic order accompanies the
breakdown of the heavy Fermi liquid at the QCP. An important question in this context
has been whether the nature of magnetic order is local or itinerant close to
the QCP where different materials appear to provide
different indication.\cite{rmp07} Of particular interest are materials where the heavy quasiparticles
seem to lose their integrity across the QCP. \cite{si01}   In this case, it is challenging
to understand how magnetism can emerge out of the heavy Fermi liquid, since
no conventional mechanism such as a spin-density wave instability is
applicable.\cite{rmp07}

In order to concentrate on the role of Kondo screening 
across the transition, theoretical concepts have been discussed that ignore 
long-range magnetic fluctuations at the quantum phase transition, such that
the local moments enter a spin liquid state upon tuning the system
across the QCP.
Although this {\em fractionalized} Fermi liquid has been shown to be a stable
phase of matter,\cite{senthil03,senthil04} in heavy Fermion materials, it was long believed
that ordered ground states are more natural to occur. Recent measurements
on chemically substituted \YbRhSi\ suggest a different situation.\cite{friedemann09} 
Upon partially replacing Rh by Ir,
magnetic order and the heavy Fermi liquid state appear to separate and a non-Fermi
liquid state without any apparent magnetic order extends
over a finite range of magnetic field strength. Similar observations
of non-Fermi liquid behavior without indication for a magnetic QCP have been reported in other 
Yb-based three-dimensional heavy fermion systems, including Ge-substituted \YbRhSi\ \cite{custers10} and
$\beta$-YbAlB$_4$.\cite{nakatsuji08} Another relevant material 
is the geometrically frustrated Kondo lattice Pr$_2$Ir$_2$O$_7$.\cite{nakatsuji06}

In this paper, we consider a fractionalized Fermi liquid state as proposed by Senthil 
et al. as a possible candidate state for such phases.\cite{senthil03,senthil04}
There, the  Kondo-Heisenberg 
lattice model is used as its microscopic starting
point. In a large-N treatment, such a model naturally describes
the breakdown of Kondo screening at a QCP where the heavy Fermi liquid is destroyed.
This breakdown is described by the hybridization mean field between local moments
and conduction electrons. Within a large-N theory, the hybridization amplitude vanishes
at the QCP, as well as the Kondo temperature derived from the hybridization mean field.\cite{senthil04}
Upon including gauge field fluctuations into this model, the transition
becomes a deconfining transition of a U(1) gauge field such that the local moment system 
forms deconfined spinon excitations. The resultant state has been coined 
fractionalized Fermi liquid, abbreviated by FL$^\ast$.
The theory shows several appealing features
that are consistent with experimental observations.\cite{senthil04,paul08} One particular
result is a logarithmic divergence of the specific heat coefficient
at the QCP, as measured in \CeAu.\cite{loehneysen96} 
Another prediction is a jump of both electrical conductivity
and Hall coefficient across the QCP.\cite{coleman05} The latter 
has been inferred from measurements 
on {\YbRhSi}  compounds,\cite{friedemann10} while differing interpretations 
of the findings in these materials are also possible.\cite{hackl10}
Whereas several subsequent works have analyzed properties of this model at the quantum critical 
point in great detail,\cite{pepin08,paul08prl,paul08,coleman05} not
much attention has been assigned to explicit predictions of thermodynamic
properties of the FL$^\ast$ phase that are measurable in experiment. Although recent experimental efforts have 
promised novel possibilities to explore such phases in materials based on 
the parent compound \YbRhSi,\cite{friedemann09,custers10} it still remains difficult to identify
such phases in terms of concrete measurements: in addition to the fact that 
spin liquid states do not break any lattice symmetries,
fermionic particles in a U(1) spin liquid state also obey the kinematic constraint of not coupling to the 
external electrical field. The existence of such degrees of freedom hence cannot be inferred
from their contribution to the electrical conductivity or the Hall
effect.

The determination of measurable signatures of a fractionalized Fermi
liquid is the main task of our paper. The angle from which we intend
to tackle this problem is that despite of not carrying electrical charge,
fermionic spinons are still able to carry heat and to respond to external 
magnetic fields. These properties can be used to analyze signatures of fractionalized
phases and help to either substantiate its existence or non-existence in forthcoming experiments. In this paper, we discuss
the thermal conductivity and the spin susceptibility
for the fractionalized Fermi liquid candidate state proposed in Ref.~\onlinecite{senthil04}.
Our results are based on a low energy effective theory for a U(1) fermionic
spin liquid state. As shown previously,\cite{senthil03} the fermionic spinon excitations
do not count to the Fermi volume and can be treated independently
from the conduction electron degrees of freedom. The stability of such
a state beyond the mean filed limit is still
under debate in two spatial dimensions, while its existence is established
in $d\geq3$ which is the main focus of the paper: building up on
previous works for the two-dimensional case, we particularly
investigate the three-dimensional setup and compare with previous
results for the 2d case. 

Starting from the low-energy effective theory, we derive the Keldysh equations
of motion for the non-equilibrium spinon Green's functions in presence of
a thermal gradient. We linearize these equations of motion and argue
that they can be rewritten in form of a variational principle that
allows to derive the thermal conductivity from a suitable variational ansatz
for the non-equilibrium spinon propagator. Collision processes
in our description originate both from impurities and collisions with U(1)
gauge bosons. In a temperature regime above the impurity-dominated regime of the thermal conductivity,
we argue that the spinons give rise to the dominant contribution to the heat current
as compared to the heat current carried by conduction electrons,
phonons, or gauge bosons. We will explain that the spinon contribution leads to a characteristic maximum in the temperature 
dependence of the Wiedemann-Franz ratio at a temperature that is strongly dependent
on the impurity concentration in the sample.

Our discussion of the spin susceptibility is based on the free energy 
contribution of the combined system of spinons and gauge bosons. Based
on this quantity, we obtain that the temperature dependence of the spinons
is logarithmically enhanced as compared to the temperature dependence
of conduction electrons in a Fermi liquid scenario. We find the zero temperature paramagnetic susceptibility 
contribution of the conduction electrons and spinons to be larger than
the conduction electron Pauli susceptibility and thus the low-temperature susceptibility promises to be 
a suitable quantity to identify fermionic spinon excitations. 




The theory of a fractionalized Fermi liquid has been originally
developed in the context of unconventional QCPs in
heavy fermion materials.\cite{senthil03,senthil04} Lateron, several more detailed
discussions of the quantum critical regime associated with this transition 
appeared.\cite{paul08,pepin08,kim09} In particular, predictions have been made
for a diverging Gr\"uneisen ratio and a violation of the 
Wiedemann-Franz ratio.\cite{kim08,kim09} 
Furthermore, spatially inhomogeneous
solutions in the heavy Fermi liquid phase have been discussed,\cite{paul08prl} 
and it has been predicted that volume collapse
transitions are likely to occur near such quantum phase transitions in sufficiently
soft materials.\cite{hackl08} Neither such first order transitions nor inhomogeneous
solutions in the Kondo screened phase could be uniquely related
to the scenario within the FL$^\ast$ phase. Instead, we attempt to
approach the problem directly from the FL$^\ast$ trial state and derive
its signatures for some directly measurable observables.
 
It is frequently stated
that the fermionic U(1) spin liquid has a stable low-energy fixed
point in the limit of large number of fermion flavors. In two dimensions,
the existence of such a limit has been questioned recently,\cite{sungsik09} due to the existence
of infinitely many corrections to the low energy fermion propagator
from planar diagrams. The subsequent proposal of an expansion in the Fermi surface
genus has been challenged by Metlitski and
Sachdev.\cite{metlitski10} 
In our work we will not elaborate on the existence of a large-N limit and just take the
previous 2d
results as valid and given, assuming that the corrections to the
one loop results in 2d are small for finite N. In three spatial dimensions
there exist no singular 
corrections from higher loop contributions which is why we can
disregard these problems in our calculations. 

Since self-energy corrections due to the fluctuating gauge field are singular, 
transport properties cannot be described by assuming well-defined quasiparticles.
Kim et al. circumvented this problem by formulating a quantum Boltzmann equation
(QBE) for a generalized distribution function.\cite{kim95} This approach has been implemented 
in form of a variational principle by Lee and Nave in order
to discuss thermal transport properties in two
dimensions.\cite{nave07} 
Properties of the magnetic susceptibility of two-dimensional fermions coupled
to a gauge field have been discussed by Lee and Nave.\cite{nave07b} 
Similarly, Ioffe and Kalmeyer have discussed these issues for bosons coupled
to a gauge field.\cite{ioffe91} Building up on these previous works, we will in particular compare the contributions from the
spinons and conduction electrons to thermal conductivity and magnetic
susceptibility, and mainly discuss three-dimensional candidate materials.


The remainder of the paper is organized as follows.
In Section \ref{sec:mod}, we introduce
some technical details of the low energy effective theory we will
use to describe the FL$^\ast$ phase.
In Section~\ref{sec:gen}, we discuss several basic concepts 
that are required for a transport theory involving 
the effective low energy description.
In Section~\ref{sec:qbe}, we use these aspects to formulate a 
quantum Boltzmann equation approach for systems in presence
of thermal gradients and derive a solution to this 
transport equation by using a variational principle. Furthermore, we discuss 
the different contributions to the total thermal conductivity in different
temperature regimes. In Section~\ref{sec:susc}, we discuss properties of the spin susceptibility
stemming from its spinon contribution. We finally conclude in
Section~\ref{summary} that heat transport and magnetic susceptibility
are promising measurements to reveal thermodynamic signatures of the fractional
Fermi liquid. 


\section{Model}
\label{sec:mod}

\begin{figure}
\includegraphics[width=7.0cm]{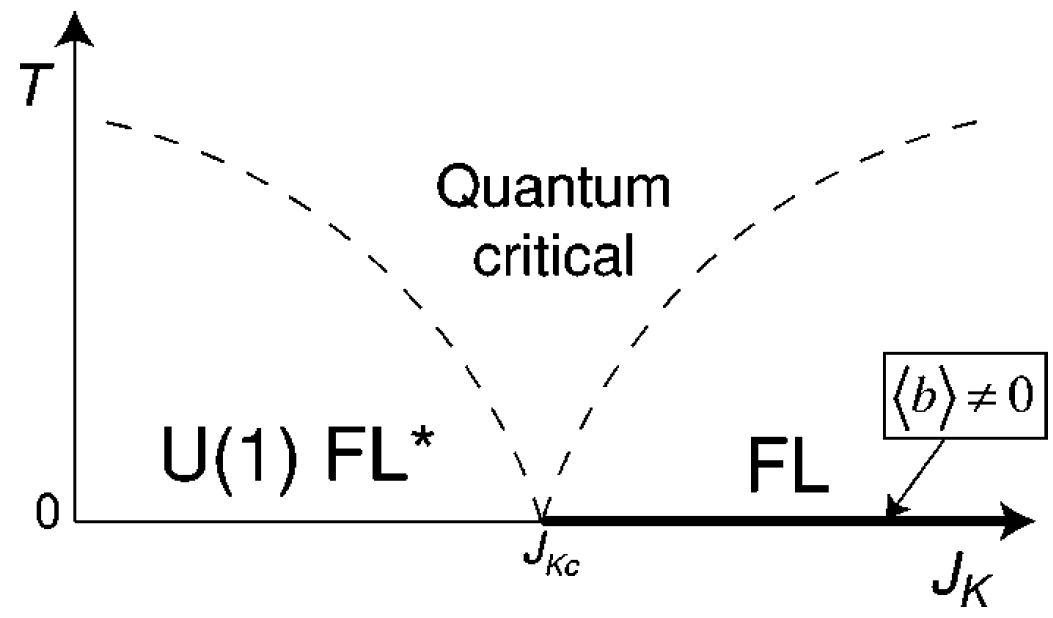}
\caption{
\label{crossovers}
Crossover phase diagram for the Kondo breakdown transition. In this paper, we analyze 
transport properties in the $U(1)$ FL$^\ast$ phase. Upon increasing temperature
in this phase, transport properties are expected to show a crossover due to the 
fluctuating bosonic hybridization field $b_i^\dagger$ in the quantum
critical region. Figure taken from Ref.~\onlinecite{senthil04}.
}
\end{figure}

A microscopic starting point for an FL$^\ast$ phase is a Kondo-Heisenberg model,
both in two and three spatial dimensions. Besides the limits of one or
infinite spatial dimensions,\cite{si01} a detailed solution of this model 
is still out of reach, and the existence of a fractionalized phase in this
model has been only established in the limit of large-N.
Very recently, an application of the ADS-CFT correspondence has 
been used to describe such FL$^\ast$ phases in a novel way, starting from the
more general Anderson lattice model.\cite{sachdev10} In this work, we will not add to the discussion of microscopic
origins of the FL$^\ast$ phases, but concentrate rather on its thermodynamic properties.
We start with the fermionic local moment representation
\begin{equation}
S_i^a=\sum_{\alpha,\beta=1}^Nf_{i\alpha}^\dagger\biggl( \Gamma_{\alpha\beta}^a\biggr)f_{i\beta}^\pdag \ ,
\end{equation}
where $\Gamma^a$ are the generators of the $SU(N)$ group in the fundamental representation.
The fermions $f_{i\alpha}^\pdag$ obey canonical anticommutation relations as well
as the local constraint  
\begin{equation}
\sum_{\alpha} f_{i\alpha}^\dagger f_{i\alpha}^\pdag=\frac{N}{2} \quad \forall \quad i \ .
\end{equation}
This constraint is enforced by a U(1) gauge field which can be minimally coupled to the spinons. 
In all following calculations, we will set $N=2$, with only slight modifications 
required for general values of $N$.
In general, the fermionic degree of freedom can be complicated by its coupling to the conduction
electrons, described by the bosonic hybridization field $b_i^\dagger=\sum_{\sigma}f_{i\sigma}^\dagger c_{i\sigma}^\pdag$. 
We will call the particle excitations of this field ``hybridization bosons" in the following.
At zero temperature, the fractionalized Fermi liquid state is reached technically by tuning the chemical
potential of the hybridization boson $b_i$ according to $\mu_b \sim J_K -J_{Kc}$, see Fig.~\ref{crossovers}.
The Kondo coupling $J_K$ itself can be tuned by variables as pressure or chemical composition.
The size of the mass gap $\mu_b$ is related to a crossover temperature scale T$^\ast$
for the FL$^\ast$ phase. In three dimensions, $T^{\ast}$ follows
$T^\ast \sim|\mu_b|^{2/3}\sim|J_K-J_{K_c}|^{2/3}$. Below temperatures of order T$^\ast$, 
the number of hybridization bosons is exponentially small as a function of temperature, and 
the system of spinons is then described by a U(1) gauge theory which is 
decoupled from the conduction electrons. In this case to which we will
constrain our attention, the effective 
action assumes the form
\begin{eqnarray}
S&=&S_c+S_f +S_a\nonumber\\
S_c&=&\int_0^\beta d\tau \sum_{{\bf k}} \bar{c}_{{\bf k}\sigma}(\partial_\tau -\varepsilon_{\bf k})c_{{\bf k}\sigma} \nonumber\\
S_f&=& \int_0^\beta d\tau \sum_r \bigl[ \bar{f}_{r\sigma}(\partial_\tau-ia_{\tau}-\mu)f_{r\sigma}\nonumber\\
   &&+\frac{1}{2m_f} \bar{f}_{r\sigma} \bigl( -i \partial_i- a_i\bigr)^2f_{r\sigma}\bigr] \nonumber\\
S_a&=& \sum_{{\bf q},\omega_n} D(q,\omega_n)^{-1} \bigl( \delta_{ij}-\frac{q_iq_j}{q^2} \bigr)a_i({\bf q},\omega_n) a_j(-{\bf q},-\omega_n) \ . \nonumber\\
\label{flast_action}
\end{eqnarray}
The conduction electron degrees freedom $c_{{\bf k}\sigma}$ are entirely described
by their excitation energies $\epsilon_{\bf k}$. The fermionic spinons $f_{{\bf k}\sigma}$
are coupled to a U(1) gauge field with transverse modes $a_i$ and longitudinal mode $a_{\tau}$.
The gauge field dynamics is entirely generated by the spinon matter field,
with the retarded propagator of the transverse fluctuations at long wavelengths
given by $D(q,\omega)=(i\Gamma(\omega/q)+\chi_dq^2)^{-1}$. Here, the dynamics
of the longitudinal gauge field mode $a_\tau$ is neglected, since its fluctuations are
short-ranged. Assuming that the spinons have mass $m_f$ and Fermi wavevector $k_F$,
the parameters $\Gamma=\frac{\pi}{k_F}$ and
$\chi_d=\frac{1}{m_f}\frac{1}{k_F^2}$ yield in $3d$, while 
we have $\chi_d=\frac{1}{12\pi m_f}$ and $\Gamma =k_F$ in $2d$. 
The action $S_f +S_a$ has been discussed in many different contexts.\cite{nagaosarmp}
Here, we consider it as a starting point in order to analyze physical observables of the FL$^\ast$
phase.


\section{General transport properties}
\label{sec:gen}

\subsection{Linear response}

We are interested in the response of the total electrical current density ${\bf J}$ and the 
total heat current density ${\bf Q}$ to an applied electrical field ${\bf E}$ and a temperature gradient
${\vec \nabla}T$, weak enough in order to linearize ${\bf J}$ and ${\bf Q}$ in the applied fields.
These properties may be formulated individually for each type of particle,
which we label by $\alpha=b,c,f$ for bosons, conduction electrons and spinons, respectively.
The transport coefficients for these individual types of particles are defined by the relations
\begin{eqnarray}
{\bf J}_{\alpha}&=&\kappa_{0\alpha} ({\bf E}_\alpha -\vec{\nabla} \mu_\alpha)+\kappa_{1\alpha}\biggl[\frac{-\vec{\nabla} T}{T} \biggr] \nonumber\\
{\bf Q}_{\alpha}&=&\kappa_{1\alpha} ({\bf E}_\alpha -\vec{\nabla} \mu_\alpha)+\kappa_{2\alpha} \biggl[\frac{-\vec{\nabla} T}{T} \biggr] \ ,
\label{linearresp}
\end{eqnarray}
where the Onsager relation for thermoelectrical transport coefficients has been used.\cite{ziman60}
The field coupling to the electrical transport coefficient $\kappa_{0\alpha}$ can originate both
from an electrical field or a gradient in the chemical potential.
Precise definitions for the current densities ${\bf J}$ and ${\bf Q}$ will appear in concrete calculations in Section \ref{sec:qbe}.
The thermal conductivity $\kappa$ is defined by the relation ${\bf Q}=-\kappa \vec{\nabla} T$ with the boundary condition
${\bf J}=0$. This implies that in general, the total thermal conductivity 
$\kappa$ will depend on all of the coefficients entering
Eq.~\eqref{linearresp}. This complicated dependence can be slightly simplified
by rewriting the thermal conductivity in form of a variational functional as 
described by Ziman,\cite{ziman60} such that not all
of the $9$ coefficients $\kappa_{\{0,1,2\}\alpha}$ need to be evaluated explicitly. We will introduce this
approach in explicit calculations later on. Moreover, any concrete calculation becomes simplified
considerably if the contribution of conduction electron and spinon degrees of
freedom to the total thermal conductivity can be calculated separately.
This case is not justified in general, since Eq.~\eqref{linearresp} couples 
different electrical fields ${\bf E}_{\alpha}$ to each type of particle.
More precisely, corrections to the approximation of independent spinon and conduction electron 
thermal currents arise for temperatures above or the same size as the crossover temperature T$^\ast$
related to the boson mass gap $\mu_b$.
A formal discussion of transport properties in this general 
situation can be given in terms of the composition rules first 
discussed by Ioffe and Larkin.\cite{ioffe89,pepin08} 

\subsection{Ioffe-Larkin rules}

The fact that the boson variable $b_i^\dagger$ is given by $b_i^\dagger = \sum_\sigma f_{i\sigma}^\dagger c_{i\sigma}^\pdag$
immediately implies the relation $\mu_b=\mu_f-\mu_c$ between the chemical potentials of boson and fermions, 
while the electrical fields felt by the particles are 
\begin{equation}
{\bf E}_b={\bf e}-{\bf E} \qquad {\bf E}_f={\bf e} \qquad {\bf E}_c={\bf E} \ .
\end{equation}
Here, the electrical field ${\bf e}$ is the fictitious field strength felt by the gauge charge of a spinon
in presence of a finite external electrical field ${\bf E}$. Since both bosons and spinons couple to the
internal gauge field ${\bf a}$, the gauge current densities will fulfill the 
constraint ${\bf J}_b+{\bf J}_f=0$. The physical electrical current density is given by ${\bf J}={\bf J}_c+{\bf J}_b$.
From these constraints, one can readily derive the composition rules stated in Ref.~\onlinecite{kim09},
representing extensions to the composition rules of Ioffe and Larkin.\cite{ioffe89}
For the total thermal conductivity $\kappa_t$, these rules imply the form\cite{kim09}
\begin{eqnarray}
\frac{\kappa_t}{T}=\frac{\kappa_{2c}}{T}+\frac{\kappa_{2f}}{T}+\frac{\kappa_{2b}}{T}
-\frac{(\kappa_{1b}+\kappa_{1f})^2}{\kappa_{0b}+\kappa_{0f}}-\frac{\kappa_{1t}^2}{\kappa_{0t}} \ .
\end{eqnarray}
Here, $\kappa_{0t}$ is the total electrical conductivity and $\kappa_{1t}$ the total 
thermoelectric conductivity. 
The correction terms 
$\frac{\kappa_{2b}}{T}-\frac{(\kappa_{1b}+\kappa_{1f})^2}{\kappa_{0b}+\kappa_{0f}}$ 
become exponentially small as a function of temperature below the crossover temperature scale 
following $T^\ast\sim|\mu_b|^{2/3}$ in three dimensions,\cite{senthil04}
since the number of bosons decreases exponentially as function of temperature in this regime.
In the same way, $\kappa_{0t}$ and $\kappa_{1t}$ reduce to the 
conduction electron contributions $\kappa_{0c}$ and $\kappa_{1c}$, since
the spinons carry no physical electrical charge. It might be possible instead to measure
the spin conductivity of the spinons, which shows a temperature dependence identical to the gauge charge  
conductivity. Its temperature-dependence behaves as $\sim T^{5/3}$ in three 
spatial dimensions.\cite{kim10} We do not elaborate on the
spin conductivity as it is a rather complicated quantity to be
measured experimentally.

Above the crossover scale T$^\ast$, the presence of hybridization bosons provides both a strong source of scattering
and a sizable channel for heat conduction, and we cannot neglect their contribution
to transport observables. Throughout the following calculations, we assume that $T<T^\ast$
and neglect therefore the presence of the hybridization boson degrees of freedom.
Therefore, we can assume that the conduction electron system shows
conventional Fermi liquid properties,
while the spinon system will be responsible for non-Fermi liquid properties.

\subsection{Spinon self-energy} 
\label{sec:spinself}

In order to analyze scattering processes of spinons on gauge field
fluctuations, it is convenient to consider the transverse gauge, $\nabla \cdot {\bf a}=0$.
In this case, the scalar ($a_0$) and transverse parts of the gauge field fluctuations
decouple. $a_0$ corresponds to a density-density response
function and does not lead to singular corrections to the spinon dynamics.\cite{nagaosarmp}
It is therefore possible to concentrate on the transverse fluctuations.
The spinon self-energy due two single gauge boson exchange processes has been 
discussed by many authors.\cite{gan93,nagaosarmp} Close to the spinon Fermi surface, the spinon self-energy corrections
are
\begin{equation}
\Sigma_a({\bf k},\omega)=\lambda_1 |\omega|^{2/3}\text{sgn}(\omega) +i\lambda_2 |\omega|^{2/3}
\label{eq7}
\end{equation}
in $2d$ and
\begin{equation}
\Sigma_a({\bf k},\omega)=\lambda_1 \omega \ln\biggl(\frac{\Lambda}{\omega}\biggr)+i\lambda_2 \omega
\label{eq8}
\end{equation}
in $3d$.\cite{gan93} Here, $\lambda_1$ and $\lambda_2$ are constants that depend on details of
the spinon Fermi surface.
Due to these self-energy corrections, the quasiparticle lifetime is not much longer
than its inverse energy and several non-Fermi liquid properties result both in
two and in three spatial dimensions.\cite{gan93} An additional complication arises 
at finite temperatures, since  $\Sigma_a({\bf k},\omega)$
is divergent in $d\leq3$ for $\omega<T$.\cite{nagaosarmp} 
In order to describe transport properties of the coupled system of spinons 
and gauge field, Kim et al. formulated a quantum Boltzmann 
equation approach in which the low energy fluctuations of the gauge field
are split off from the collision integral in order to avoid the divergent
self-energy.\cite{kim95}\\ 
To cure the finite temperature divergence of the single-particle self-energy, vertex corrections
are of central importance. The divergence mentioned above is caused by the fact that the single-particle
spinon Green's function is not invariant under a U(1) gauge transformation.\cite{nagaosarmp}
In order to calculate a gauge invariant quantity such as thermal 
conductivity, it is also possible to include the low-energy gauge
fluctuations into a quantum Boltzmann equation, since the divergences are expected to cancel in the 
gauge invariant conductivities. This situation occurs in our following
calculations.

At low frequencies the spinons are also influenced by 
impurity scattering for which we assume potential scattering with potential strength $V({\bf r})$.
Any type of impurity with short-ranged Coulomb potential will lead to such scattering processes.
We assume in the following that the scattering potential is independent of temperature,
as well as a temperature independent concentration and a Gaussian distribution 
of impurities in the bulk of the sample. Modifications to 
the impurity potential are expected to occur only at temperatures of order the ionization 
energy required to change the impurity charge, which is considerably higher than room temperature. 
To lowest order in the impurity concentration $n_i$, such impurity scattering processes are described 
by\cite{mahan90} 
\begin{equation}
\Sigma_{\text{imp}}({\bf k},\omega)=n_iT_{{\bf k}{\bf k}}
\end{equation}
with the $T$-matrix given by the solution to
\begin{equation}
T_{{\bf k}{\bf k}^\prime}(\omega)=V({\bf k}-{\bf k}^\prime)+
\int \frac{d^3{\bf p}}{(2\pi)^3}\frac{V({\bf k}-{\bf p})T_{{\bf p}{\bf k}^\prime}(\omega)}{\omega-\xi_{\bf p}+i\delta} \ .
\end{equation}
Here, $\xi_{\bf p}={\bf p}^2/(2m_f)-\mu$ is the spinon dispersion and $V({\bf p})$ is the Fourier
transform of $V({\bf r})$.
In the following, we neglect interference processes between scattering on gauge bosons and 
impurities, such that the total self-energy is given by 
$\Sigma({\bf k},\omega)=\Sigma_{\text{imp}}({\bf k},\omega)+\Sigma_a({\bf k},\omega)$. 
This approximation is valid if the mean free path derived from one of these two collision processes
is much larger than that of the second collision process. Since the mean free path is
given by the spinon Fermi velocity times the spinon lifetime $v_F/\Sigma({\bf k}_{\bf F},\omega)$, it is sufficient 
to compare the size of the two self-energy contributions $\Sigma_{\text{imp}}({\bf k},\omega)$ 
and $\Sigma_a({\bf k},\omega)$ for this purpose. Setting them equal defines a characteristic temperature 
which we denote by $T_1$.
First, the neglect of interference processes is therefore valid
in the temperature regime $T_1\ll T<T^\ast$ where scattering on gauge field fluctuations dominates, such that
$\Sigma({\bf k},\omega) \simeq \Sigma_a({\bf k},\omega)$. The temperature scale $T^\ast$
is the crossover temperature for the FL$^\ast$ phase boundary we introduced above.
Second, our approximation is also valid for temperatures $T\ll T_1$ where impurity scattering dominates
the self energy such that $\Sigma({\bf k},\omega) \simeq n_iT_{{\bf k}_{\bf F}{\bf k}_{\bf F}} (\omega=0)$. 
The temperature scale $T_1$ is set by equating the imaginary parts 
$\Sigma^{\prime\prime}_{\text{imp}}({\bf k}_{\bf F},\omega=0)=\Sigma^{\prime\prime}_a({\bf k}_{\bf F},\omega)$
and considering $\hbar\omega\simeq k_BT$. We abbreviate the impurity self-energy as 
the spinon scattering rate $\tau_f^{-1}$ such that $\Sigma_{\text{imp}}({\bf k}_{\bf F},\omega=0)=\tau_f^{-1}$.
For the constant $\lambda_2$ in Eqs~\eqref{eq7} and~\eqref{eq8}, we use the estimate
$\lambda_2\simeq (\epsilon_F^f)^{1/3}$ in 2d.\cite{blok93} In 3d, an estimate for metals
is given by $\lambda_2 \simeq 10^{-5}$,\cite{kwon95} although the gauge field coupling of the spinons 
can differ to some extend from the case of electrons in a metal.
In this way, we obtain
\begin{equation}
k_BT_1\simeq \biggl[\frac{\hbar}{\tau_f (\epsilon_F^f)^{\frac{1}{3}}} \biggr]^{3/2} 
\end{equation}
in $2d$ and
\begin{equation}
k_BT_1\simeq 10^5\frac{\hbar}{\tau_f}  
\end{equation}
in $3d$. We thus conclude that the temperature window where the neglect of interference processes is not justified is proportional 
to $\tau_f^{(d-5)/2} \sim n_i^{(5-d)/2}$ and therefore shrinks rapidly with decreasing impurity concentration.
Concise numbers for the temperature $T_1$ can obtained by estimating $\tau_f$ 
from the residual conductivity expression~\eqref{cdrude}, which itself could
be estimated from the conduction electron residual conductivity.
Another way to check the justification of the neglect of interference processes is
to compare the mean spacing between impurities and the scattering length on gauge bosons.
The latter will shrink as a function of increasing temperature and can thereby become much smaller
than the mean spacing between impurities.

\section{Quantum Boltzmann equation}
\label{sec:qbe}

For a fermionic quantum many-body system under non-equilibrium 
conditions, we define the usual Keldysh propagators \cite{mahan90}
\begin{eqnarray}
G^>(x_1,x_2)&=&-i\langle\psi(x_1)\psi^\dagger(x_2) \rangle \nonumber\\
G^<(x_1,x_2)&=& i\langle\psi^\dagger(x_1)\psi^\pdag(x_2) \rangle\nonumber\\
G_t(x_1,x_2)&=&  \theta(t_1-t_2)G^>(x_1,x_2) \nonumber\\
            &+&  \theta(t_2-t_1)G^<(x_1,x_2)\nonumber\\
G_{\bar{t}}(x_1,x_2)&=& \theta(t_2-t_1)G^>(x_1,x_2)\nonumber\\
                    &+& \theta(t_1-t_2)G^<(x_1,x_2)
\end{eqnarray}
and formulate a Dyson equation
\begin{equation}
\tilde{G}=\tilde{G}_0+\tilde{G}_0 \tilde{\Sigma}\tilde{G}
\label{dyson}
\end{equation}
for the matrix Green's function
\begin{equation}
\tilde{G}=
\left[
\begin{array}{cc}
G_t  & - G_< \\
G_>  &  -G_{\bar{t}}\\
\end{array}
\right]
\label{mg}
\end{equation}
and $\tilde{\Sigma}$ defined in analogy. $\tilde{G}_0$ contains the non-interacting propagators.
We employed the combined variable $x=({\bf r},t)$
that parametrizes the fermionic field $\psi$. Without further complications,
it can also include spin and other variables.
For brevity, we dropped these coordinates in Eqs~\eqref{dyson} and~\eqref{mg} as well 
as we dropped integrating over intermediate coordinates in Eq.~\eqref{dyson}.\\
For transport processes, it is convenient to introduce
a center-of-mass coordinate system 
\begin{eqnarray}
({\bf R},T)&=& \frac{1}{2}(x_1+x_2) \nonumber\\
({\bf r},t)&=& x_1-x_2 \ .
\end{eqnarray}
Starting from the Dyson equation~\eqref{dyson} it is then
possible to derive an equation 
of motion for $G^<(x_1,x_2)$, which we refer to as {\em quantum Boltzmann equation}.
This has been discussed in detail by Mahan.\cite{mahan90} For these
purposes, it is convenient to work in the Fourier space of relative
coordinates ${\bf r}$
\begin{equation}
\tilde{G}({\bf k},\omega,{\bf R},T)
=\int d^d{\bf r}e^{i {\bf k} {\bf r}} \int d t  e^{i\omega t}  \tilde{G}({\bf r},t,{\bf R},T) \ .
\end{equation}

\subsection{Derivation}

Our transport description starts from the gauge invariant quantum Boltzmann equation~\cite{mahan90} 
\begin{eqnarray}
&i&\biggl\{ \frac{\partial}{\partial T} + {\bf v}_{\bf k} \cdot \nabla_{\bf R} 
+e{\bf E} \cdot\biggl[ \biggl(1-\frac{\partial \text{Re}\Sigma_{\text{ret}}}{\partial \omega}
\biggr)\nabla_{\bf k} \nonumber\\
&+& ({\bf v}_{\bf k}+\nabla_{\bf k}\text{Re}\Sigma_{\text{ret}})\frac{\partial}{\partial \omega} \biggr] \biggr\}G^< 
-ie{\bf E} \cdot\biggl[ \frac{\partial \Sigma^<}{\partial \omega} \nabla_{\bf k} \text{Re} G_{\text{ret}}  \nonumber\\
&-&\frac{\partial \text{Re}G_{\text{ret}}}{\partial \omega} \nabla_{\bf k} \Sigma^<\biggr] 
=\Sigma^>G^< -\Sigma^<G^> \nonumber\\
&+&i[\text{Re}\bigl[\Sigma_{\text{ret}}\bigr],G^<] + i[\Sigma^<,\text{Re}\bigl[G_{\text{ret}}\bigr]] \ .
\label{fullmotion}
\end{eqnarray}
Here, we employed the Poisson bracket
\begin{eqnarray}
[\Sigma,G]&=&\frac{\partial \Sigma}{\partial \omega} \frac{\partial G}{\partial T} 
-\frac{\partial \Sigma}{\partial T} \frac{\partial G}{\partial \omega} \nonumber\\
&-& \nabla_{\bf k} \Sigma \cdot \nabla_{\bf R} G + \nabla_{\bf R} \Sigma \cdot \nabla_{\bf k} G 
\label{poissonbrack}
\end{eqnarray}
and used the retarded propagator $G_{\text{ret}}$ defined by the relation $G_{\text{ret}}=G_t-G^<$.
Further simplifications to Eq.~\eqref{fullmotion} can be made in absence of an external electrical 
field (${\bf E}=0$), such that all terms proportional to ${\bf E}$ can be dropped. 
In addition we linearize both hand sides in $\nabla_{\bf R}$. 
Here, we just state the result that we justify further in Appendix~\ref{solveqbe},
\begin{equation}
A \frac{\partial f}{\partial \omega} \frac{\nabla_{\bf R}T}{T}\cdot {\bf v}_{\bf k}\omega 
=\Sigma^>G^<-\Sigma^<G^> \ .
\label{qbefinal}
\end{equation}
In the driving terms, the quantities $\Gamma=-2\text{Im}\Sigma_{\text{ret}}$, $A=i(G^>-G^<)$
and the Fermi function $f(\omega)=(1+\exp(\beta\omega))^{-1}$, $\beta=1/k_{\text{B}}T$,
appear. From the solution to this equation, the thermal current density is obtained as
\begin{equation}
{\bf Q} = -i\int\frac{d^d{\bf k}}{(2\pi)^3}\frac{{\bf k}}{m_f}\int \frac{d\omega}{2\pi}\omega G^<({\bf k},\omega) \ .
\end{equation}
By linearizing this equation in the driving field $\nabla_{\bf R}T$,
it is possible to obtain the thermal conductivity $\kappa$
if the boundary condition ${\bf J}=0$ is obeyed.\\
In order to specify the collision term in Eq.~\eqref{qbefinal}, we consider the 
non-equilibrium self-energy contributions from single gauge boson exchange processes
\begin{eqnarray}
&&\Sigma_{\text{a}}^<({\bf k},\omega)=\sum_{\bf q} \int_0^\infty \frac{d\nu}{\pi} \biggl| \frac{{\bf k}\times {\bf{\hat q}}}{m_f}\biggr|^2
\text{Im}D({\bf q},\nu)\biggl[\{n(\nu)+1\} \nonumber\\
&&G^<({\bf k}+ {\bf q} ,\omega+\nu) +n(\nu)G^<({\bf k}+ {\bf q} ,\omega-\nu)\biggr] \ ,
\label{gaugeboson}
\end{eqnarray}
and the contribution from potential scattering
\begin{equation}
\Sigma_{\text{imp}}^<({\bf k},\omega)=n_i\int \frac{d^3{\bf p}}{(2\pi)^3} [T_{{\bf k}{\bf p}}(\omega)]^2G^<({\bf p},\omega) \ .
\end{equation}
The total self-energy is given by the sum $\Sigma^<=\Sigma_{\text{a}}^<+\Sigma_{\text{imp}}^<$.
We note that in this approximation, we have approximated the gauge field propagator 
by its equilibrium form.  
This approximation implies that the presence of a 
thermal current of gauge bosons does not change the spinon thermal conductivity.
No contradiction with previous approximations arises.
Although without impurities, the system has the translational symmetries
of the Bravais lattice, the finite spinon thermal conductivity we will obtain 
within the approximation of an equilibrated gauge field is not artificial. A well-known analogy is
the electrical conductivity of a Fermi liquid, which has a well-established temperature
dependence that can be obtained from scattering due to quasiparticle interactions,
without involving mechanisms that break translational invariance.
We will comment further on this approximation when we discuss the thermal 
conductivity of the gauge bosons.

\subsection{Variational principle}

It turns out that Equation~\eqref{qbefinal} is too complicated to solve in this form.
We attempt to simplify it further such that it can be rewritten in form
of a variational principle for the distribution 
function $G^<({\bf k},\omega)$. We make the linearized ansatz
\begin{equation}
G^<({\bf k},\omega)=iA ({\bf k},\omega)\biggl[f(\omega) 
- \frac{\partial f(\omega)}{\partial \omega} \frac{\nabla_{\bf R} T}{T} \cdot {\bf v}_{\bf k} 
\Lambda({\bf k},\omega)\biggr] \ ,
\label{linearized}
\end{equation}
which is valid in the regime $\bigl|\nabla_{\bf R} T / T \bigr|\ll 1$. All non-equilibrium properties are encoded
in the vertex distribution function $\Lambda({\bf k},\omega)$ which
still needs to be determined.
First, we assume that the collision term is dominated by scattering on the gauge field,
$\Sigma_a^< \gg \Sigma_{\text{imp}}^<$. This approximation is valid 
in the temperature regime $T_1 \ll T<T^\ast$ discussed above.
Employing Eqs~\eqref{gaugeboson} and~\eqref{linearized}, the collision term can be specified as
\begin{eqnarray}
&&i[\Sigma^< A-2\Gamma G^<] = 2\Gamma ({\bf k},\omega)A({\bf k},\omega) \frac{\partial f}{\partial \omega}
\frac{\nabla_{\bf R} T}{T} \cdot {\bf v}_{\bf k} \Lambda({\bf k},\omega)\nonumber\\
&+&A({\bf k},\omega) \sum_{\bf q} \int_0^\infty \frac{d\nu}{\pi} 
\biggl|\frac{{\bf k} \times \hat{{\bf q}}}{m_f} \biggr|^2 \text{Im} D({\bf q},\nu) 
\biggl[ \frac{\nabla_{\bf R} T}{T} \cdot {\bf v}_{{\bf k}+{\bf q}} \biggr]
\nonumber\\
&&\frac{\partial f(\omega)}{\partial \omega}\biggl\{[n(\nu)+f(\omega+\nu)]\frac{1-f(\omega+\nu)}{1-f(\omega)}\nonumber\\
&&A({\bf k}+{\bf q},\omega+\nu)\Lambda({\bf k}+{\bf q},\omega+\nu) -[n(-\nu)+f(\omega-\nu)]\nonumber\\
&&\frac{1-f(\omega-\nu)}{1-f(\omega)}A({\bf k}+{\bf q},\omega-\nu)\Lambda({\bf k}+{\bf q},\omega-\nu)] \biggl\} \ .
\end{eqnarray}
Solving this equation for the vertex distribution function yields the self-consistency condition
\begin{eqnarray}
&&\Lambda({\bf k},\omega)=\frac{\omega}{2\Gamma({\bf k},\omega)} -\frac{1}{2\Gamma({\bf k},\omega)} \sum_{\bf q} 
\int_0^\infty \frac{d\nu}{\pi}
\biggl| \frac{{\bf k}\times{\hat{{\bf q}}}}{m_f}\biggr|^2 \nonumber\\
&&\text{Im}D({\bf q},\nu) \frac{{\bf v}_{{\bf k}+{\bf q}}\cdot{\bf v}_{{\bf k}} }{v_{\bf k}^2} 
\biggl\{ [n(\nu)+f(\omega+\nu)]\frac{1-f(\omega+\nu)}{1-f(\omega)} \nonumber\\
&& A({\bf k}+{\bf q},\omega+\nu)\Lambda({\bf k}+{\bf q},\omega+\nu)- [n(-\nu)+f(\omega-\nu)] \nonumber\\
&&\frac{1-f(\omega-\nu)}{1-f(\omega)}A({\bf k}+{\bf q},\omega-\nu) \Lambda({\bf k}+{\bf q},\omega-\nu) \biggr\} \ .
\label{vertexdist}
\end{eqnarray}
Equation~\ref{vertexdist}
is solved by a vertex distribution function $\Lambda(\omega)$ 
that depends only on frequency but not on momentum. This result
is of central importance in order to formulate a variational 
ansatz for $G^<({\bf k},\omega)$ later on.
To begin with, $\Gamma({\bf k},\omega)$ does not depend
on momentum according to Eqs~\eqref{eq7} and~\eqref{eq8},
such that $A({\bf k}+{\bf q},\omega)$ depends on ${\bf k}$ only through the variable 
$\xi=\epsilon_{{\bf k}+{\bf q}}-\mu$. The collision integral is dominated
by gauge boson frequencies $\nu \simeq q^3 \chi_d/\Gamma$, with the  
frequency being set by $\nu \simeq k_BT$. In the limit $k_BT \ll k_F^3 \chi_d/\Gamma \simeq \epsilon_F^f$,
we can therefore approximate $q\ll k_F$, such that all other terms in the integrand 
of Eq.~\eqref{vertexdist} approximately depend only on $q$ but not on ${\bf k}$,
with corrections of higher order in the small parameter $q$. We will give an estimate for 
the spinon Fermi energy $\epsilon_F^f$ below.
Within this approximation, the integral can be entirely parametrized 
in terms of the variables $\xi=\epsilon_{\bf k+q}-\mu$
and $q$, and the value of the integral will not depend on ${\bf k}$. Therefore, $G^<({\bf k},\omega)$
will depend on the variable $\xi=\epsilon_{\bf k}-\mu$ only through the equilibrium 
spectral function $A({\bf k},\omega)$, which is sharply peaked as a function of $\xi$. 
This variable can therefore be eliminated from the quantum Boltzmann equation, and we 
can make use of the reduced distribution function 
\begin{equation}
f(\hat{{\bf k}},\omega)=-i\int \frac{d\xi}{2\pi}G^<(\xi,\hat{{\bf k}},\omega) \ .
\end{equation}
In this case, it is possible to represent 
the thermal conductivity as a variational 
functional.\cite{ziman60} A similar approach has been employed previously by
Nave and Lee \cite{nave07} in the case of two dimensions. We repeat their
steps wherever it is instructive in order to understand our results
for the three dimensional case.

Introducing an ansatz of the form $f(\hat{{\bf k}},\omega)=f(\omega)-\phi(\hat{{\bf k}},\omega)\frac{\partial f}{\partial \omega}$
immediately leads to  
\begin{equation}
\phi(\hat{\bf k},\omega) \sim \frac{\nabla_{\bf R}T}{T} \cdot {\bf v}_{{\bf k}} \Lambda(\omega) \ .
\end{equation}
Assuming a power-law dependence $\Lambda(\omega) \propto \omega^s$,\cite{foot3} the trial function
is given by $\phi(\hat{{\bf k}},\omega) = \omega \hat{{\bf k}} (\nabla_{\bf R} T/T)$. 
Upon inserting the definitions for $\dot{S}$ and ${\bf Q}$ into the general form of 
Eq.~\eqref{variation}, it is seen that the power $s$ in the ansatz 
$\Lambda(\omega) \propto \omega^s$ drops out of the temperature dependence of the 
thermal conductivity once frequency is rescaled with temperature. Therefore, 
we dropped the exponent $s$ from our variational ansatz.
Note that it is not necessary to normalize $\phi(\hat{{\bf k}},\omega)$ since
the normalization constant will drop out of our final results.
We next define the rate of entropy density production as
\begin{equation}
\dot{S}=\vec{\nabla}\biggl(\frac{1}{T}\biggr)\cdot {\bf Q}
\end{equation}
generated by the thermal current density ${\bf Q}$.
$\phi$ minimizes the variational functional 
\begin{equation}
W=\frac{T^2 \dot{S}}{Q^2} \ ,
\label{variation}
\end{equation}
where ${\bf Q}$ is the heat current density given by 
\begin{equation}
{\bf Q}=N(0)\int d \hat{\bf k} d\omega \phi v_F \omega \frac{\partial f_0}{\partial \omega} {\bf \hat{k}}
\end{equation}
and the rate $\dot{S}$ of entropy density production is
\begin{eqnarray}
\dot{S}&=&\frac{N_0^2}{m_f^2}\beta^2\int d\omega d\omega^\prime d\nu d {\bf q} d \hat{{\bf k}} d\hat{{\bf k}}^\prime 
|{\bf k}^\prime \times \hat{{\bf q}}|^2 \nonumber\\
&\times& {\text{Im}} D({\bf q},\nu)(\phi-\phi^\prime)^2 f_0(\omega)[1-f_0(\omega^\prime)]n_0(\nu)  \nonumber\\
&\times& \delta(\omega^\prime-\omega-\nu)\delta((\hat{{\bf k}}^\prime-\hat{\bf k})-\frac{\bf q}{k_F} ) \ .
\label{rateofentropy}
\end{eqnarray}
Here the integration over ${\bf \hat{k}}$ is
defined as integration over the $d-1$ dimensional unit sphere,
$N(0)$ is the density of states at the Fermi level for up and down spins combined.
In $2d$, we have $N(0)=m_f/\pi$ and in 3d $N(0)=m_fk_F/(2\pi^2)$
The global minimum of the variational functional $W$ is identical to the spinon thermal 
resistivity $\kappa_f^{-1}$.\\
As discussed in Section~\ref{sec:spinself}, at lowest temperatures the energy relaxation rate is impurity dominated
\begin{equation}
\Gamma ({\bf k},\omega) \sim 2\Gamma_0 \ ,
\end{equation}
assuming a constant impurity contribution $2\Gamma_0=2n_i\text{Im} \bigl[T_{{\bf k_F}{\bf k_F}}(\omega=0) \bigr]$.
In this case, the rate of entropy density production in Eq.~\eqref{rateofentropy} has to be modified accordingly.
Instead, this case can be treated in a much simpler way by using the Wiedemann-Franz law for free fermions
\begin{equation}
\frac{\kappa_f}{\kappa_{0f}T}=L_0 \ ,
\label{wfratio}
\end{equation}
with the Lorentz number $L_0=(\pi^2/3) (k_B/e)^2$.
We can use the Drude formula for the residual conductivity of
a Fermi gas
\begin{equation}
\kappa_{0f}= \frac{n_f e^2\tau_f}{m_f} 
\label{cdrude}
\end{equation}
and insert it into Eq.~\eqref{wfratio} to determine the spinon heat conductivity $\kappa_{f}$. 
The spinon density $n_f$ is equal to the inverse unit cell volume, while  
the spinon scattering rate $\tau_f^{-1}$ can be equated to twice the self-energy correction
from impurity scattering, $2\Gamma_0$. 
Independent of dimension, up to a number of $\mathcal{O}(1)$,
the heat current density is given by $Q\equiv N(0) v_F(k_BT)^2$.\\
In order to determine $\dot{S}$ in Eq.~\eqref{rateofentropy},
we first integrate over the variables $\omega,\omega^\prime$ and ${\bf q}$ and parametrize
the integrals over $\hat{{\bf k}}$ and $\hat{{\bf k}}^\prime$ in spherical coordinates.
Furthermore, we can approximate
\begin{equation}
(\phi-\phi^\prime)^2=\nu^2({ \hat{\bf k} \cdot \hat{\bf u}})^2 + 2\nu(\omega -\mu)^2 \mathcal{O}(q) \ ,  
\end{equation}
where $\nu$ is given by the relative frequency $\nu=\omega-\omega^\prime$.
Since the main contribution to the integral comes from $q\sim\nu^{1/3}$,
the term $\nu^2({ \hat{\bf k} \cdot \hat{\bf u}})^2$ dominates the temperature
dependence  at $k_BT \ll \epsilon_F^f$ and we can neglect the term $2\nu(\omega -\mu)^2 \mathcal{O}(q)$ 
and all higher corrections. A straightforward but somewhat tedious integration 
shows then that the rate of entropy production follows $\dot{S} \sim (k_BT)^{(6+d)/3}$.\\
We note that we did not introduce an infrared cutoff frequency for the gauge field
mode in Eq.~\eqref{rateofentropy}, although this leads to a divergent retarded self-energy at
finite temperatures.\cite{nagaosarmp} Even at one loop order, this divergence is canceled
in the collision integral of the QBE. Introducing an infrared cutoff given by $\beta \nu=1$ 
would only renormalize the prefactor of the overall result by a number of $\mathcal{O}(1)$, 
as we checked explicitly.

\subsection{Thermal conductivity: results}

Putting together the Drude result and the result for 
scattering on gauge bosons, the spinon conductivity is given by
\begin{eqnarray}
\frac{\kappa_f}{T}=\left\{
\begin{array}{cc}
 \biggl[ \frac{\hbar}{k_B^2}\bigl(\frac{k_BT}{\epsilon_F^f}\bigr)^{2/3} +\frac{3}{\pi^2} \frac{m_f}{k_B^2n_f \tau_f} \biggr]^{-1} \quad (2d) \nonumber\\
 \biggl[\frac{\hbar}{k_B^2}\frac{k_BT}{k_F\epsilon_F^f}  +\frac{3}{\pi^2} \frac{m_f}{k_B^2n_f \tau_f} \biggr]^{-1} ~\qquad \quad  (3d)
\end{array}
\right\} \ . \nonumber\\
\label{kresult}
\end{eqnarray}
Here, we restored units of $\hbar$ and $k_B$ and omitted prefactors of $\mathcal{O}(1)$.
The spinon density $n_f$ corresponds simply to the inverse of the unit cell volume, such that
$n_f=1/V_{uc}$.
We defined the unit cell volume $V_{uc}$ as the volume of the two-dimensional unit cell
for the two-dimensional thermal conductivity and as the volume of the three-dimensional unit cell for the three-dimensional thermal
conductivity. Therefore, we defined the two-dimensional thermal conductivity as the thermal conductivity of a single 
two-dimensional layer.
It turns out from the results in Eq.~\eqref{kresult} that in both two and three dimensions, 
the spinon thermal conductivity  agrees with the basic argument
$\kappa \sim C_V v_F^2 \tau_E$ stated by Nave and Lee,\cite{nave07} assuming that
the specific heat follows $C_V \sim \gamma T$
and $\tau_E$ is the energy relaxation rate of the spinons, which is constant
for isotropic impurity scattering and follows $\tau_E \sim
T^{-d/3}$ 
for scattering on gauge field fluctuations.\cite{nagaosarmp} 
The result for the two-dimensional thermal conductivity has been already obtained by Nave and Lee.\cite{nave07}
For the case of three-dimensional candidate materials, we have
now obtained the finding
that the thermal conductivity of the spinons
due to scattering on gauge field fluctuations has a temperature-
independent contribution. This can be seen from the result for 3d in Eq.~\eqref{kresult},
since $\kappa_f/T \sim 1/T$ in the limit $\tau_f^{-1} =0$.
However, due to impurity scattering the total thermal conductivity will vanish 
in the limit of zero temperature, such that the constant term $d_2$
appearing in Table~\ref{table1} can only appear above the temperature scale $T_1$
introduced in \ref{table1}.
\begin{table}[]
\begin{tabular*}{6.8cm}{|c|c|c|c|}
\hline
\hline
&(I) $T \ll T_1$ & (II) $T_1 \ll T<T_2$ & (III) $T_2<T$ \\
\hline
2d & $c_1 T$ & $c_2 T^{1/3}$         & non-universal  \\
3d & $d_1 T$ &  $d_2$ & non-universal  \\
\hline
\hline
\end{tabular*}
\caption{\label{table1}
Temperature-dependence of the total thermal conductivity $\kappa$
in different temperature regimes. 
Estimates for the characteristic temperatures $T_1$ and $T_2$
as well as expressions for the parameters $c_1$, $c_2$, $d_1$
and $d_2$ are discussed in the main text. The temperature-dependence
in regime (III) is non-universal since it depends also on
the temperature scale $T^\ast$, among other aspects. It is discussed in more detail in the main text.}
\end{table} 
It is important to recall that we have approximated the collision integral
by the sum of impurity and gauge field collision processes, and it has been
assumed that any interference between these types of collision processes does not influence the
temperature dependence of the spinon thermal conductivity.

Finally, the contribution of conduction electrons to the thermal conductivity
can be obtained by using the Wiedemann-Franz law~\eqref{wfratio},
\begin{equation}
\kappa_c=L_0 \rho^{-1} T ,
\end{equation}
with the Fermi liquid resistivity \cite{coleman}
\begin{eqnarray}
\rho&=&\rho_0+a \frac{h}{2e^2} \biggl(\frac{\Gamma}{2\epsilon_F^c}\biggr) \nonumber\\
\Gamma&=&\frac{1}{2}(\pi k_BT)^2/\epsilon_F^c \ .
\end{eqnarray}
Here, $a$ is the interlayer spacing of the metallic system, $\epsilon_F^c$
the conduction electron Fermi energy and the residual resistivity $\rho_0$ can be 
determined from the Drude formula~\eqref{cdrude}.
In metallic systems, the Lorentz number $L_0$ is up to a factor of $\mathcal{O}(1)$ 
equivalent to the free electron value used in~\eqref{wfratio}.

We now summarize the behavior of the total thermal conductivity $\kappa$
given by
\begin{equation}
\kappa=\kappa_c+\kappa_f \ .
\end{equation}
Based on our results, we distinguish the quantitative behavior of three different temperature
regimes (I)-(III) summarized in Table~\ref{table1}.\\
Regime (I) is the impurity dominated regime with a linear temperature dependence
as in a conventional Fermi liquid.
The ratio of spinon vs. conduction electron contribution in this regime is 
\begin{equation}
\frac{\kappa_f}{\kappa_c}=\frac{m_c}{m_f} \frac{n_c\tau_c}{n_f \tau_f} \ .
\end{equation}
$m_c/m_f$ is a small factor that is approximately given by the 
ratio $J_H/D$ of the spinon dispersion to conduction electron band width $D$.
Here, $J_H$ is understood as the strength of nearest neighbor RKKY exchange that is naturally   small
compared against the conduction electron band width.
This could yield to a dominance of the conduction electron contribution to the
impurity dominated thermal conductivity, although
the relaxation times also play a role in this ratio.\\
Regime (II) is the non-Fermi liquid regime dominated by the temperature
dependence of the spinon contribution which extends from $T_1$ to some
larger temperature scale $T_2$. It shows a rather sluggish $T^{1/3}$ dependence in two dimensions 
and a constant contribution in three dimensions.
The temperature-dependence in regime (III) depends on various degrees of freedom,
such that the temperature scale $T_2$ depends not only on parameters of the 
spinon system. First, the leading temperature dependence of the spinon thermal 
conductivity in regime (II) is only valid at temperatures $k_BT< \epsilon_F^f$,
as we explained in the context of equation~\eqref{vertexdist}. Next, 
spinon scattering on hybridization bosons can only be neglected below the
temperature scale $T^\ast$. We note that also gauge bosons carry a
finite thermal current that leads to corrections to the total thermal conductivity.
As discussed below, this correction is again negligible for temperatures $k_B T<\epsilon_F^f$.
We can estimate the spinon Fermi energy $\epsilon_F^f$ by equating it to the typical energy scale $J_H$
for RKKY exchange interaction, $\epsilon_F^f\approx J_H\approx 100K$.
Finally, the temperature dependence of the conduction electron 
contribution has to be compared against the temperature dependence
of the spinon thermal conductivity. We use the estimate
\begin{equation}
\biggl[\frac{\epsilon_F^c}{k_BT}\biggr]^2=\biggl[\frac{\epsilon_F^f}{k_BT} \biggr]^{\frac{d}{3}}  
\end{equation} 
for the temperature scale when both contributions become equal
and obtain $k_BT=\epsilon_F^c \bigl( \epsilon_F^f/\epsilon_F^c\bigr)^{(d-1)/2}$.
Since $\bigl( \epsilon_F^f/\epsilon_F^c\bigr)\ll1$, this temperature scale
is much larger than the conduction electron Fermi energy, which is of order
$10^4$K, and we can therefore neglect the temperature dependence of the 
conduction electron thermal conductivity in this case.
We conclude that the temperature scale $T_2$ is given by the minimum of
the energy scales $\epsilon_F^f$ and $T^\ast$.\\ 
A more detailed discussion of regime (III) relies on additional high energy degrees of freedom
that we excluded from our theory so far, and a more detailed analysis of this temperature
regime is required to accurately discuss the thermal conductivity beyond the temperature scale $T_2$.

\subsection{Wiedemann-Franz ratio}

A   suitable quantity to detect spinon contributions to the thermal conductivity 
is the Wiedemann-Franz ratio 
\begin{equation}
L(T)=\frac{\kappa_t}{\sigma_t T}  \ .
\end{equation}
Due to the absence of a spinon charge current, the presence of a spinon thermal current
will always imply $L(T)>\kappa_c/(\sigma_cT)$. 
Since in our theory the conduction electron Wiedemann-Franz ratio $\kappa_c/(\sigma_cT)$ behaves as in 
a Fermi liquid, the additional spinon contribution will lead to characteristic deviations
from Fermi liquid behavior. In the following,
we distinguish two characteristic
regimes for the Wiedemann-Franz ratio that are identical to the 
regimes (I) and (II) for the thermal conductivity of the last subsection.\\
In regime (I), the impurity dominated regime, the Wiedemann-Franz ratio
is given by 
\begin{equation}
\frac{\kappa_t}{\sigma_t T}=L_0(1+\frac{u_f}{u_c})
\end{equation}
with $u_f=(n_f\tau_f)/m_f$ and $u_c=(n_c\tau_c)/m_c$. 
The absence of a spinon electrical current causes a renormalization 
of the Wiedemann-Franz ratio in the impurity dominated regime
that might be unobservably small, due to the small mass ratio $m_c/m_f \sim J_H/D\ll 1$.\\
\begin{figure}
\includegraphics[width=7.0cm]{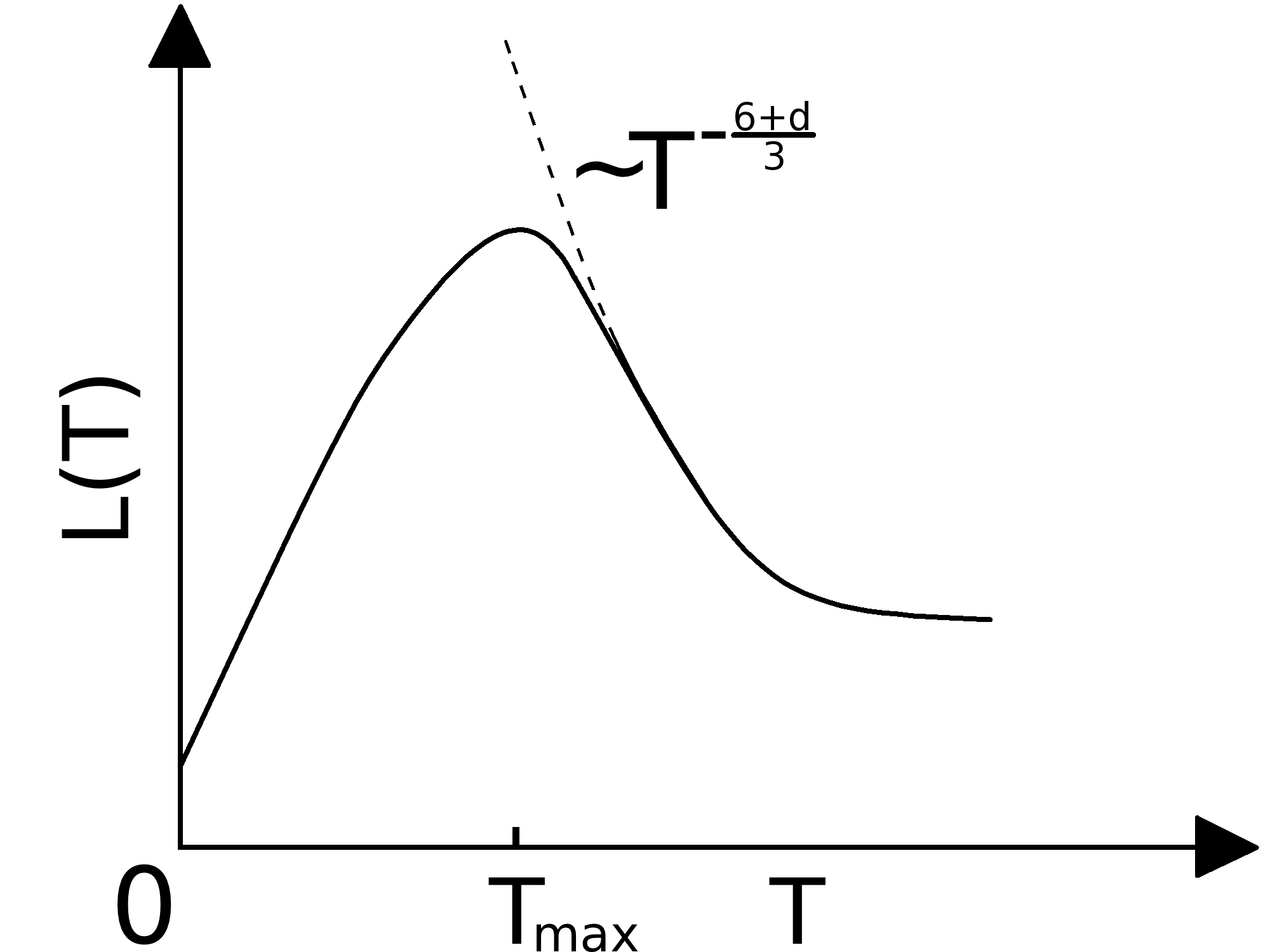}
\caption{
\label{wfpic}
Sketch of the characteristic temperature dependence of the Wiedemann-Franz ratio $L(T)$
in the temperature regimes (I) and (II) discussed in the main text.
The quantity $L(T)$ corresponds to the full line.
At zero temperature, the ratio saturates at the finite value $L(T=0)=L_0(1+\frac{u_f}{u_c})$.
At a temperature $T_{max}$ that is i.e. dependent on the spinon and conduction electron scattering rate,
a maximum occurs that turns over into a characteristic power law divergence $T^{-(6+d)/3}$
that is sketched in form of the dashed line. 
}
\end{figure} 
More promising with respect to observability is regime (II), where spinon collisions are dominated by scattering on gauge bosons.
There, the Wiedemann-Franz ratio is given by the ratio of temperature-dependent 
parts of spinon thermal conductivity 
and conduction electron electrical conductivity. 
In this case the Wiedemann-Franz ratio shows the temperature dependence
\begin{eqnarray}
L(T)\sim\left\{
\begin{array}{cc}
  (k_BT)^{-\frac{8}{3}}  \quad (2d) \nonumber\\
  (k_BT)^{-3}  \quad  (3d)
\end{array}
\right\} \ . \nonumber\\
\label{wfarr}
\end{eqnarray}
These divergences will be cut off at low temperatures by the thermal conductivity 
due to impurity scattering.
Therefore, the behavior stated in Eq.~\eqref{wfarr} leads to a characteristic maximum of the Wiedemann-Franz ratio
at a temperature $T_{max}$ that can be estimated by equating the two contributions to Eq.~\eqref{kresult}.
This estimate results in the temperature
\begin{equation*}
k_BT_{max}\simeq \biggl[ \frac{m_fk_F (\epsilon_F^f)^{2/3}}{\hbar n_f\tau_f} \biggr]^{\frac{3}{2}}
\end{equation*}
in two spatial dimensions and 
\begin{equation*}
k_BT_{max}\simeq \biggl[ \frac{n_fk_F \epsilon_F^f}{\hbar m_f\tau_f} \biggr]
\end{equation*}
in three spatial dimensions.
More general, any thermal conductivity with a temperature
dependence dominating the Fermi liquid $T^3$-dependence will lead
to a maximum of the Wiedemann-Franz ratio. This assumes
that the electrical conductivity of the system behaves like in a Fermi liquid.

\subsection{Corrections to thermal conductivity}
\label{sec:correct}

The contribution of gauge bosons to thermal transport can be analyzed 
in analogy to the spinon transport equation by defining a distribution 
function 
\begin{equation}
n({\bf q},{\bf r},t)=\int \frac{d\nu}{2\pi}D^<({\bf q},\nu , {\bf r},t) \ .
\end{equation}
It is now again possible to use the variational principle used already in this section
to calculate the spinon thermal conductivity if the spinons are assumed 
to be in equilibrium in the gauge boson transport equation. 
However, in principle the gauge boson thermal current 
leads to a modification of the spinon thermal current and vice versa.
This effect could be included by considering the coupled system of transport
equations for spinons and gauge bosons. In analogy to the phonon drag in metals,
such corrections to the spinon thermal conductivity can be shown to be of relative
size $(k_BT)/\epsilon_F$ and can be neglected within this accuracy.\cite{ziman60,nave07}
It is now completely analogous to the calculation of the spinon thermal 
conductivity to rewrite the gauge boson thermal conductivity in form of a variational
functional.\cite{nave07} We do not repeat these steps in detail here, but cite just
the result we obtained. After solving the integrals for the gauge boson heat current density
and entropy production, the gauge boson thermal conductivity $\kappa_g$ can be estimated as 
\begin{equation}
\kappa_g = c_g \biggl[\frac{k_BT}{\epsilon_F^f}\biggl]^{(d+1)/3} \ ,
\end{equation}
with a non-universal constant $c_g$ that depends on details of the spinon dispersion.
For temperatures $\frac{k_BT}{\epsilon_F^f}\ll 1$, the gauge boson thermal conductivity $\kappa_g$
is therefore only a subleading correction to the thermal conductivity
of the spinons, see Eq.~\eqref{kresult}.\\ 
Finally, phonon contributions to the thermal conductivity are proportional to $(T/\Theta_D)^3$
and are therefore negligible for temperatures much smaller than the Debye
temperature $\Theta_D$.

\section{Magnetic susceptibility}
\label{sec:susc}

\subsection{Results}

In our model, the physical spin susceptibility is given by the sum of the spinon susceptibility and the conduction 
electron susceptibility inside the FL$^\ast$ phase,
\begin{equation}
\chi_s=\chi_c+\chi_f \ .
\end{equation}
In analogy to our discussion of the thermal conductivity, corrections to this formula
arise due to the presence of hybridization bosons and are therefore again exponentially small 
as a function of temperature below the temperature scale T$^\ast$. We will neglect 
these effects in the following, considering again only temperatures $T<T^\ast$.
Part of the behavior of the spin susceptibility $\chi_s$ has been discussed
already by Paul et al.,\cite{paul08} with a focus mainly on the
finite temperature regime above the QCP. In the FL$^\ast$ regime
considered by us, the conduction electron contribution $\chi_c$ is 
described by the usual Fermi liquid form
\begin{eqnarray}
\chi_c=\mu_B^2\frac{2N_c(0)}{1+F_0^{a}}+\left\{
\begin{array}{cc}
b_1T ~ \quad (2d) \nonumber\\
b_2T^{2} \quad (3d)
\end{array}
\right\} \ .
\end{eqnarray}
The paramagnetic part is controlled by the conduction electron density of 
states $N_c(0)$ at the Fermi level and the Landau parameter F$_0^a$.
Different from a Fermi gas, in two dimensions linear temperature corrections arise in a Fermi liquid,\cite{chubukov05}
while in three dimensions, the usual T$^2$ dependence prevails.\cite{belitz97} The coefficients $b_1$ and $b_2$ depend 
on interactions and band structure details and while $b_1$ is positive,\cite{chubukov05} the sign of $b_2$
has not yet been found to the best of our knowledge. 
One way to obtain the spinon contribution $\chi_f$
is the free energy of the system of spinons and gauge field,\cite{foot4}
\begin{equation}
F=-k_B T \ln Z \qquad Z=\int \mathcal{D}a \int \mathcal{D}f e^{i(S_f+S_{a})} \ .
\end{equation} 
In RPA approximation, this free energy is given by $F_s+F_a$, where
$F_a$ is the correction to the free fermion part $F_s$
due to quadratic fluctuations of the transverse gauge field,
\begin{equation}
F_a=\frac{2}{\beta}\sum_{\omega_n} \int \frac{d^dk}{(2\pi)^d}\ln\biggl( \frac{\Gamma|\omega_n|}{k}+\chi_d k^2\biggr) \ .
\end{equation}
Therefore, $\chi_f$ will have the form $\chi_f=\chi_f^0+\Delta\chi_f$,
where $\chi_f^0=(g_f^2/2)\mu_B^2 N_f(0)$ has the Fermi gas form of free spinons with 
g-factor $g_f$ and density of states at the Fermi level $N_f(0)$.
The contribution $\Delta\chi_f$ denotes additional fluctuation
corrections to the susceptibility.

It yet remains to discuss the dependence of $F_a$ on an external magnetic field.
In presence of an external magnetic field, $F_a$ will split into two contributions $F_{a\sigma}$
stemming from different spin projections $\sigma=\pm \frac{1}{2}$.
The magnetic field couples
to the spinons by a Zeeman term $E_Z=g_f\mu_B B$ that splits the Fermi wavevector
to the spin-dependent values $k_{F\sigma}=\sqrt{k_F^2+\sigma 2m_f E_Z}$, depending on the
spinon g-factor $g_f$. In this way, it is straightforward
to obtain the shift $\Delta\chi_f$ from 
\begin{equation}
\Delta\chi_f=-\frac{\partial^2F_a}{\partial B^2}\biggl|_{B=0} \ .
\label{susc}
\end{equation}
The low temperature behavior is
\begin{eqnarray}
\Delta\chi_f=\left\{
\begin{array}{cc}
\chi_0 + a_1T^{\frac{5}{3}}  \qquad \qquad ~~ (2d) \nonumber\\
\chi_0 + a_2T^{2} \log(T/\epsilon_F^f) ~(3d)
\end{array}
\right\} \ .
\end{eqnarray}
Details of the calculation are shown in App.~\ref{solvesusc}, where in particular
the form of the parameters $a_1$ and $a_2$ is clarified. Both $a_1$ and $a_2$ are positive
and depend on the spinon dispersion.
The temperature-independent term $\chi_0$ is a paramagnetic contribution that
has been discussed in detail in Ref.~\onlinecite{nave07b}. 
It is roughly of the same order of magnitude as the Pauli paramagnetic susceptibility
$\chi_f^0$. 

\subsection{Discussion}

Possibly the best way to observe the spinon contribution to the susceptibility 
is to consider the zero temperature limit. In this limit, the ratio of the spinon to conduction 
electron Pauli susceptibility is given by the ratio of their densities of states
at the Fermi surface. This ratio can be estimated from the ratio of their band widths, $ D / J_H$.
The spinon band width energy scale $J_H$ is generated by RKKY exchange according to $J_H\sim J_K^2/D$, 
such that the spinon Pauli susceptibility dominates that of
the conduction electrons by a large factor $D^2/J_K^2 \gg 1$. This strong effect could therefore 
be used for the experimental identification of such an FL$^\ast$ phase with gapless fermionic spinons.
Indeed, in the non-Fermi liquid phases recently observed in experiments
on Yb-based materials, the spin susceptibility at lowest applied temperature 
was enhanced by up to a factor of three\cite{custers10} and even five\cite{friedemann09}
as compared to the Pauli susceptibility deep inside the heavy Fermi liquid phase.\\
The temperature-dependent part of the spinon susceptibility also leads
to important modifications of the full temperature dependence of the magnetic susceptibility.
In two dimensions, the asymptotic temperature dependence in the limit $T\rightarrow 0$ is set by the conduction 
electron contribution and behaves linear in temperature. The free spinon part $\chi_f^0$
does not to contribute to this asymptotic behavior, since it has the characteristic quadratic
temperature dependence of band fermions. Still, it will influence the overall temperature dependence
of the spin susceptibility. Since in general the spinon Fermi 
energy is quite small compared to the conduction electron Fermi energy, the spinon
contribution $\Delta \chi_f$ might qualitatively change the temperature dependence 
of the susceptibility at quite
low temperatures. An estimate is that the $T^{\frac{5}{3}}$-dependence due 
to gauge field corrections is dominating above a temperature scale set 
by $k_BT/\epsilon_F^c = (k_BT / \epsilon_F^f)^{5/3}$, such that 
\begin{equation}
k_B T=\epsilon_F^f \biggl(\frac{\epsilon_F^f}{\epsilon_F^c}\biggr)^{3/2} \ .
\end{equation}
Estimating $\epsilon_F^f\approx J_H \approx 100K$ and 
$\epsilon_F^c\approx 10^4 K$ we get $T \approx 0.1 K$ for the crossover to the spinon
dominated temperature-dependence of the spin susceptibility.\\
For three-dimensional candidate materials, both the temperature dependence of the gauge field contribution and
the band susceptibility of the spinons should dominate the temperature dependence of the conduction electron 
contribution to the total spin susceptibility. The contribution $\chi_f^0$ is expected to do so because 
it is proportional $(k_BT/\epsilon_F^f)^2$, with the lower Fermi energy $\epsilon_F^f\ll\epsilon_F^c$.
Ultimately, the logarithmic enhancement $\sim\log(k_BT/\epsilon_F^f)$ set by the gauge field
corrections is expected to dominate the temperature dependence of $\chi_s$ in the regime $k_BT/\epsilon_F^f \ll 1$.
In general, our results predict a positive finite temperature correction to the Pauli susceptibility.


\section{\label{summary} Summary and discussion}

We have discussed the thermal conductivity and spin susceptibility 
of a U(1) fractionalized Fermi liquid in two and three spatial dimensions.
Within the scenario of gapless fermionic spinon excitations,
the total thermal conductivity becomes enhanced by the spinon contribution
already in the impurity dominated regime, although the enhancement might be 
unobservably small due to the small Fermi velocity of the spinons. In a temperature regime
where scattering of spinons is dominated by gauge boson
absorption and emission, we were able to argue that the spinon thermal current will dominate the thermal 
conductivity as compared to conduction electron, phonon and gauge
boson contributions. For the important three-dimensional candidate state, we obtained a
characteristic plateau in the temperature dependence of the 
heat conductivity that dominates all other temperature-dependent contributions
in magnitude. We propose to detect such anomalous contributions to the 
thermal conductivity by measuring the Wiedemann-Franz ratio.
As a function of temperature, this ratio shows a pronounced temperature maximum
which is caused by the non-Fermi liquid thermal conductivity of the spinon system.
The size of the maximum depends strongly on the impurity concentration and can largely
exceed the Wiedemann-Franz ratio of conventional metals if the system
is sufficiently clean.

Our theoretical predictions particularly rely  on the existence of gapless
fermionic excitations. However, many other types of spin liquid states might 
be realized in nature with gapped/and/or bosonic excitation spectra
that do not show a Pauli paramagnetic susceptibility. 
From our analysis of the spinon susceptibility in the U(1) FL$^\ast$ phase, we predict a large enhancement of the 
Pauli paramagnetic susceptibility due to the flat dispersion 
of the fermionic spinon excitations caused by RKKY exchange. This enhancement should
be observable at temperatures $k_BT \lesssim \epsilon_F^f$.
Such a large paramagnetic contribution is unlikely to be observed in presence of bosonic
excitations or a gapped spectrum.
A gapped spinon spectrum would be 
characteristic for an increasing spin susceptibility as a function of temperature
if the size of the gap is smaller than the spinon Fermi energy.\cite{nave07} 
Such a behavior is not observed in recent experiments.\cite{friedemann09}

As an additional thermodynamic signature for our three dimensional 
candidate state, however, we predict a logarithmic enhancement of the temperature-dependent part of the Fermi 
liquid susceptibility from the spinon contribution.
Our results might be complemented by future work that may help to predict other signatures
of ordered states in the local moment system coexisting with itinerant
conduction electrons.
A   unique probe for fermionic spin liquids 
may be the thermal Hall effect.\cite{nagaosa10} Undoubtedly, this demands
further experimental effort. Moreover, microscopic calculations
need to be established that can explain the suppression of broken symmetries
in Kondo lattice systems once the heavy Fermi liquid ground state is destroyed.


\begin{acknowledgments}
We acknowledge discussions with  A. Benlagra, S.~Friedemann, L.~Fritz, P.~A. Lee, 
O.~Motrunich, A. Rosch, T. Senthil and M. Vojta.
Furthermore, AH thanks M. Vojta for collaborations on related topics.
This research was supported by the DFG through SFB~608, SFB-TR/12, and FG 960.
AH is supported by the David and Ellen Lee Foundation. RT is
supported by a Feodor Lynen scholarship of the Humboldt Foundation and
Alfred P. Sloan Foundation funds.
\end{acknowledgments}


\appendix

\section{\label{solveqbe} Derivation of quantum Boltzmann equation}

It is possible to rewrite the collision term in~\eqref{qbefinal} as
\begin{equation}
\Sigma^> G^< -\Sigma^<G^>=-i [2\Gamma G^< -\Sigma^< A ] \ .
\label{appcollision}
\end{equation}
According to our assumption, the gauge field propagator is replaced by
its equilibrium form, and the drag between spinons and gauge field
is neglected. Furthermore, the retarded spinon propagator obeys the equation of motion\cite{mahan90}
\begin{equation}
\biggr[\omega-\epsilon_{\bf k}+\frac{1}{8m_f} \nabla_R^2-\Sigma_{\text{ret}} \biggl]G_{\text{ret}}=1 \ .
\end{equation}
In the linearized spatially homogeneous quantum Boltzmann equation,
the retarded self-energy $\Sigma_{\text{ret}}({\bf k},\omega)$ can therefore be replaced by its equilibrium form,
since field strength enters only to $\mathcal{O}({\bf E}^2)$ or $\mathcal{O}(\nabla_{\bf R}^2)$.
Since the collision term~\eqref{appcollision} has to vanish in equilibrium,
only field dependent terms have to be retained, and the collision term reads
therefore
\begin{eqnarray}
&-&i [2\Gamma G^< -\Sigma^< A]   = 2\Gamma (k,\omega)A(k,\omega) \biggl[ \frac{\partial f}{\partial \omega}\biggr] 
{\bf F}\cdot {\bf v}_{\bf k} \Lambda({\bf k},\omega)\nonumber\\
&+&A(k,\omega) \biggl[ \sum_q \int_0^\infty \frac{d\nu}{\pi} \biggl|\frac{k \times \hat{q}}{m_f} \biggr|^2 \text{Im} D({\bf q},\nu) \biggl[{\bf F}\cdot {\bf v}_{{\bf k}+{\bf q}} \biggr]
\nonumber\\
&&\biggl[ \frac{\partial f(\omega)}{\partial \omega}\biggr]\biggl\{[n(\nu)+f(\omega+\nu)]\frac{1-f(\omega+\nu)}{1-f(\omega)}\nonumber\\
&&A(k+q,\omega+\nu)\Lambda(k+q,\omega+\nu) -[n(-\nu)+f(\omega-\nu)]\nonumber\\
&&\frac{1-f(\omega-\nu)}{1-f(\omega)}A(k+q,\omega-\nu)\Lambda(k+q,\omega-\nu)] \biggl\} \ .
\end{eqnarray}
The driving force ${\bf F}$ is given by the ratio $\nabla T/T$ in our concrete case.
It remains to derive the driving term of the QBE~\eqref{qbefinal}.\\
In presence of a thermal gradient, we can still neglect any dependence of the
QBE on the absolute time variable $T$.
However, the spatially dependent temperature distribution causes a
dependence of the functions $\Sigma^<$ and $G^<$ on the coordinate ${\bf R}$.
The driving term therefore includes the terms
\begin{eqnarray}
&-&\nabla_{\bf R} \text{Re}\Sigma_{\text{ret}} \cdot \nabla_{\bf k}G^< + \nabla_{\bf k}(\epsilon_{\bf k}+\text{Re}\Sigma_{\text{ret}})\cdot \nabla_{\bf r}G^<\nonumber\\
&-&[\Sigma^<,\text{Re}G_{\text{ret}}] \ .
\end{eqnarray}
We can use the arguments of Ref.~\onlinecite{nave07} in order to show that the term
$[\Sigma^<,\text{Re}G_{\text{ret}}]$ has no influence on physical observables. Moreover, the gradients
$\nabla_{\bf R} \text{Re}\Sigma_{\text{ret}}$ and $\nabla_{\bf k} \text{Re}\Sigma_{\text{ret}}$
vanish according to the assumptions stated above.
Therefore, the driving term is simply given by
$\nabla_{\bf k} \epsilon_{\bf k} \nabla_{\bf r}G^<$.\\

\section{\label{solvesusc} Calculation of susceptibility}

Following the same steps as in Ref.~\onlinecite{senthil04}, we obtain for the temperature-dependent part
of the free energy
\begin{eqnarray}
F(T)-F(0)&=& \frac{4}{\pi} \int \frac{d^dk}{(2\pi)^d}\int_0^\infty du e^{-u \chi_dk^3} \frac{1}{u^2} \nonumber\\
&\times&\biggl[\frac{\pi}{2}u\beta^{-1}\coth\biggl( \pi\frac{u\Gamma}{\beta}\biggr)-\frac{1}{2\Gamma}\biggr] \ .
\end{eqnarray}
The effect of a magnetic field can be analyzed afterwards by
considering the field-dependence of the parameters $\Gamma$ and $\chi_d$.\\
We introduce the new variables $x=\frac{2Tu}{q^3}$ and $u=\frac{\Omega}{2T}$
and get in $2d$
\begin{eqnarray}
F(T)-F(0)&=& T^{\frac{5}{3}}\biggl(\frac{\Gamma}{\chi_d}\biggr)^{\frac{2}{3}}\frac{\Gamma(\frac{2}{3},0)}{3\pi^2}13.3425 +\mathcal{O}(T^2)\ .
\end{eqnarray}
In three dimensions, we get 
\begin{eqnarray}
F(T)-F(0)=T^2\biggl( \frac{\Gamma}{\chi_d}\frac{1}{9\pi} \log \biggl( \frac{\chi_d\Lambda^3}{\Gamma T}\biggr) \biggr) +\mathcal{O}(T^2)\ ,
\end{eqnarray}
with a momentum cutoff $\Lambda$ that is of order the spinon Fermi wavevector $k_F$.
The quantity $\chi_d \Lambda^3/\Gamma$ is equivalent to the spinon Fermi energy up to 
a factor of $2/\pi$.
Here and in the following, we employ units where $k_B=1$.
In a finite magnetic field, the ratio $\frac{\Gamma_\sigma}{\chi_{d\sigma}}$ depends on the spin projection $\sigma$, 
\begin{eqnarray}
\frac{\Gamma_\sigma}{\chi_{d\sigma}}=\left\{
\begin{array}{cc}
12\pi k_{F\sigma} m_f \quad (2d) \nonumber\\
  \pi k_{F\sigma} m_f \qquad (3d)
\end{array} \right\} \ ,
\end{eqnarray}
with the shifted Fermi wavevectors $k_{F\sigma}=\sqrt{k_F^2 +\sigma g_f 2 m_f\mu_B B}$.
The temperature-dependent part of the susceptibility ~\eqref{susc} is finally obtained 
by using
\begin{eqnarray}
-\frac{d^2\bigl(\frac{\Gamma}{\chi_d}\bigr)^{\frac{2}{3}}}{dB^2} &=&
\frac{5}{36} k_F^{-\frac{7}{3}} (12\pi m_f)^{\frac{2}{3}} (g_f m_f \mu_B)^2 \quad (2d)\nonumber\\
-\frac{d^2\bigl(\frac{\Gamma}{\chi_d}\bigr)}{dB^2} &=&  \frac{\pi}{4}\frac{m_f}{k_F^3} (g_f m_f \mu_B)^2\quad (3d) \ .
\end{eqnarray}
The temperature-independent contribution to Eq.~\eqref{susc} can be obtained in a similar way.\cite{nave07}



\begin{thebibliography}{51}
\bibitem{foot3} The vertex distribution function $\Lambda(\omega)$ encodes the scattering
processes that stabilize a stationary thermal current. In the impurity
dominated regime, it can be replaced by a constant.
Since the one loop self-energy correction of the spinons has a power law dependence on frequency
both in 2d and 3d, 
it is expected that $\Lambda(\omega)$ can be described by a power law in the regime where
scattering process on gauge field fluctuations dominate.
\bibitem{foot4}
The same results could be also obtained from the 
spin-spin correlation function dressed with the 
gauge field self-energy corrections.
\end{thebibliography}
\end{document}